\documentclass{article}

\PassOptionsToPackage{numbers,sort&compress}{natbib}
\usepackage[dblblindworkshop,final]{neurips_2026}
\workshoptitle{SaTQuML: Secure and Trustworthy Quantum Machine Learning}

\usepackage[utf8]{inputenc}
\usepackage[T1]{fontenc}

\usepackage{amsmath,amssymb,amsfonts,amsthm}
\usepackage{booktabs}
\usepackage{graphicx}
\usepackage{hyperref}
\usepackage{microtype}
\usepackage{multirow}
\usepackage{nicefrac}
\usepackage{xcolor}
\hypersetup{
  hidelinks,
  pdftitle={Where Quantum Fourier Sampling Stops Short: A Three-Gate Audit Protocol for Delay-PUF Security Models},
  pdfauthor={Owen Friedewald; Ali Shiri Sichani; Chi-Ren Shyu}
}

\graphicspath{{figures/}}
\newtheorem{proposition}{Proposition}
\newcommand{\F}{\mathbb{F}}

\newcommand{\NS}{\operatorname{NS}}
\newcommand{\sgn}{\operatorname{sgn}}

\title{Where Quantum Fourier Sampling Stops Short:\\
A Three-Gate Audit Protocol for Delay-PUF Security Models}
\author{%
  \href{https://orcid.org/0009-0003-6847-1706}{Owen Friedewald}\quad
  \href{https://orcid.org/0000-0002-7239-1341}{Ali Shiri Sichani}\quad
  \href{https://orcid.org/0000-0001-9197-9522}{Chi-Ren Shyu}\\
  Department of Electrical Engineering and Computer Science\\
  University of Missouri, Columbia, MO, USA\\
  \texttt{omfvq4@missouri.edu}\quad
  \texttt{asp9f@missouri.edu}\quad
  \texttt{shyuc@missouri.edu}
}

\begin{document}
\maketitle

\begin{abstract}
Quantum Fourier sampling (QFS) appears to offer a direct route to auditing the
spectral learnability of delay-based physical unclonable functions (PUFs).  We
ask whether that promise survives access matching, a strong classical
comparator, and oracle synthesis, and report where it stops short.  Three gates
structure the evaluation.  \textbf{Structure:} low degree is not small support
at reachable challenge lengths; for 4-XOR at $n=14$, degree
$\le d_f(0.1)$ admits $91\%$ of all $2^n$ characters and the median
$90\%$-mass set spans a third of the spectrum.  \textbf{Algorithmics:}
constructing the phase oracle logically implies classical membership access,
making Kushilevitz--Mansour the correct baseline; across 45 tasks it exhausts
each finite domain, and no 4-XOR ideal-sampling case reaches
$90\%$ mass within $2^n$ calls.  A quantum-kernel diagnostic appears more
favorable, with geometric difference rising to $2.151$ at $N=512$ challenges, but it
correlates $0.991$ with $1/\sqrt{\lambda_{\min}(K_C)}$ for the classical Gram
matrix $K_C$, and the 4-XOR
label-complexity ratio does not exceed a balance-preserving permutation null
($p=0.930$).  Trace-normalized geometric difference can therefore grow through
classical ill-conditioning alone, without task-label alignment.
\textbf{Implementation:} a simulator-validated fixed-point phase oracle based on
the quantum Fourier transform (QFT)
admits an $18.9\%$ routed-depth reduction, yet the least certified precisions
have estimated durations of $1.18$--$1.55\times$ the median dephasing time $T_2$
of the mapped qubits
on a static backend snapshot, without hardware execution.  We find no end-to-end
advantage in the evaluated regime, although ideal sampling does use fewer
coherent calls on the thresholded task.  The contribution is the Three-Gate
Quantum Audit Protocol: a reproducible procedure separating an ideal query
advantage from a realizable security benefit.  This is not a claim about
deployed silicon and not an impossibility result.
\end{abstract}

\section{Introduction}

Delay-based physical unclonable functions (PUFs) derive device identity from
manufacturing variation rather than a stored digital secret
\citep{lee2004arbiter}.  They remain relevant as heterogeneous packages combine
dies from different supply chains: roots of trust can live in an interposer
\citep{nabeel2020root}, and between-die delay measurements can fingerprint an
assembled chiplet system \citep{deric2022chiplet}.  Their security is also a
learning problem.  Arbiter PUFs admit an additive linear delay model, and
classical modeling attacks can predict responses from challenge--response
pairs (CRPs) \citep{ruhrmair2010modeling}.  More elaborate constructions such
as XOR and interpose PUFs are designed to move that boundary
\citep{nguyen2019interpose,wisiol2020splitting}.

Quantum learning suggests a tempting audit: encode a Boolean response function
as a phase, apply Hadamards, and sample Fourier characters with probability
equal to their squared Fourier coefficients.  If an $x$-XOR arbiter PUF is
low-degree concentrated, this looks preferable to enumerating all
$\sum_{j\le d}\binom{n}{j}$ characters.  That comparison omits two costs.  The
phase oracle is a stronger access primitive than CRPs, and support enumeration
is not the strongest classical algorithm under membership-query access.

This paper asks a deliberately operational question: \emph{does the promise of
quantum Fourier sampling survive access matching, a strong classical
comparator, and implementation of the phase oracle for simulated delay-PUF
auditing?}  We use three decision criteria, which we call gates:

\begin{enumerate}
  \item \textbf{Structural:} concentration must be sparse relative to the full
  subset lattice at the measured $n$, not merely low-degree in notation.
  \item \textbf{Algorithmic:} quantum Fourier sampling must be compared with a
  classical heavy-Fourier algorithm receiving the membership access implied by
  oracle construction, and a quantum kernel must differ materially from an
  efficient classical suite before training.
  \item \textbf{Implementation:} a reversible oracle must reproduce the target
  response phase, uncompute its workspace, and preserve the downstream audit
  signal at a routed cost compatible with the target device.
\end{enumerate}

All three gates are informative, and together they locate the boundary between
an ideal query advantage and a realizable security benefit.  None supports an
end-to-end advantage in our evaluated regime.  We package them as the
\emph{Three-Gate Quantum Audit Protocol}.  Our contributions are: (i) an access-matched audit with a
Kushilevitz--Mansour (KM) comparator; (ii) a finite-$n$ evaluation of the
known XOR concentration scale relative to the full subset lattice;
(iii) an algebraic and empirical conditioning diagnosis for geometric-difference
diagnostics---in our near-identity quantum-kernel regime, $g_{CQ}$ tracks
$1/\sqrt{\lambda_{\min}(K_C)}$, while a balance-preserving permutation control
provides no evidence of 4-XOR-specific label alignment; and (iv) an exact
fixed-point phase oracle with task-level precision criteria fixed before the
sweep and resource estimates accounting for device connectivity.  The
negative result preserves what remains open: ideal Fourier sampling uses fewer
coherent calls in our finite experiments, but coherent calls and classical
model evaluations are not equivalent physical resources.

\paragraph{Relation to Fourier-analytic PUF evaluation.}
This work builds on, rather than introduces, Fourier analysis for PUF security.
The low-degree algorithm of Linial, Mansour, and Nisan (LMN)
\citep{linial1993constant} was adapted into a noise-tolerant Fourier attack and
security measure for arbiter-family PUFs by \citet{ganji2018fourier}.
Earlier PAC analysis gave algorithm-specific learnability bounds for XOR-arbiter
PUFs \citep{ganji2015attackers}; \citet[Corollary~1]{ganji2020pitfalls} then
derived the LMN scale $m=2.32x^2/\varepsilon^2$ under the uniform distribution.
PUFmeter used
noise sensitivity and property testing to evaluate resistance to modeling
\citep{ganji2019pufmeter}.  Our structural gate asks a different finite-regime
question: whether the resulting low-degree region and its mass-carrying subset
are actually small enough to support a quantum audit.  Our algorithmic and
implementation gates then add access-matched KM comparison, kernel geometry,
and a reversible-oracle cost that these classical security analyses did not
study.

\section{Setting and access model}
\label{sec:setting}

\subsection{Additive-delay PUFs}

For a challenge $c\in\{0,1\}^n$, define suffix-parity variables
\begin{equation}
u_i=\bigoplus_{j=i}^{n}c_j, \qquad \Phi_i(c)=(-1)^{u_i}.
\label{eq:suffix}
\end{equation}
The map $c\mapsto u$ is an invertible triangular linear map over $\F_2$.
Consequently it permutes Walsh characters and preserves Fourier coefficient
magnitudes.  An additive-delay arbiter PUF (APUF) is the halfspace
\begin{equation}
A(c)=\sgn\!\left(w_b+\sum_{i=1}^{n}w_i\Phi_i(c)\right),
\label{eq:apuf}
\end{equation}
with zero assigned the positive sign.  An $x$-XOR PUF is a product
$f(c)=\prod_{k=1}^{x}A_k(c)$.  We analyze the uniform challenge distribution,
for which $f(u)=\sum_{S\subseteq[n]}\widehat f(S)(-1)^{\oplus_{i\in S}u_i}$
and Parseval gives $\sum_S\widehat f(S)^2=1$
\citep{odonnell2014analysis}.

The simulator was independently cross-checked against \texttt{pypuf} v2.2.0
\citep{pypuf2020}: matched APUF, 2-XOR, and 4-XOR instances had identical
responses, quality metrics, and logistic-regression attack accuracy.  This
validates implementation parity, not the additive model as a complete model of
silicon.

\subsection{Access is part of the problem}

Table~\ref{tab:access} separates three resources that are often conflated.
Observed CRPs support ordinary supervised learning.  Classical membership
access returns $f(c)$ for a selected challenge.  Coherent access implements
$O_f\lvert u\rangle=(-1)^{[f(u)<0]}\lvert u\rangle$ on a superposition.  A
fabricated arbiter chain is a classical physical device; it has no interface
for coherent challenges.  In this work, $O_f$ can only be synthesized from a
known, simulatable delay model held by a designer.  The same model can be
evaluated classically, so oracle construction logically implies membership
access.  This implication establishes the classical comparator's access, not
equality of classical and quantum query complexity or runtime.

\begin{table}[t]
\caption{Access-matched comparison.  Query counts are meaningful within a row,
not across physically different resources.}
\label{tab:access}
\centering
\small
\begin{tabular}{p{0.22\linewidth}p{0.28\linewidth}p{0.33\linewidth}}
\toprule
Resource & Realizable holder & Required comparator \\
\midrule
Observed CRPs & passive or protocol-level attacker & tuned supervised learner \\
Classical membership $f(c)$ & chosen-challenge user or model holder & KM/Goldreich--Levin family \\
Coherent phase $O_f$ & reversible designer-held model only & Fourier sampling with full oracle cost \\
\bottomrule
\end{tabular}
\end{table}

This boundary is the most important security qualification in the paper:
\textbf{we claim no quantum attack speedup against a fabricated or deployed
PUF}.  The quantum object is a design-time model audit.

\section{Decision gates}
\label{sec:methods}

\subsection{Concentration with its denominator}

Let $d_f(\varepsilon)$ be the smallest $d$ for which
$\sum_{|S|>d}\widehat f(S)^2\le\varepsilon$.  Reporting $d_f$ alone can be
misleading: the admissible support has size
$L(n,d)=\sum_{j=0}^{d}\binom{n}{j}$, which may be nearly $2^n$.

The usual clipping argument for multiplying low-degree approximations is not
valid because clipping creates higher-degree Fourier mass.  The following
route preserves the degree statement.

\begin{proposition}[Known XOR concentration consequence]
\label{prop:xor}
Let $f=\prod_{k=1}^{x}A_k$, where each $A_k$ is a $\{-1,+1\}$-valued
halfspace under the uniform distribution.  Then
\[
  \sum_{|S|>d}\widehat f(S)^2=O\!\left(\frac{x}{\sqrt d}\right).
\]
Hence $f$ is $\varepsilon$-concentrated through degree
$d=O(x^2/\varepsilon^2)$, with formal support at most
$n^{O(x^2/\varepsilon^2)}$ when $d\ll n$.
\end{proposition}

For completeness, the proof uses the pointwise union bound
$\mathbf{1}[f(u)\ne f(u_\delta)]\le
\sum_k\mathbf{1}[A_k(u)\ne A_k(u_\delta)]$, Peres' halfspace bound
$\NS_\delta(A_k)=O(\sqrt\delta)$ \citep{peres2004noise}, and the Fourier
identity for noise sensitivity.  A complete proof is in
Appendix~\ref{app:proof}; this recovers the scale reported by
\citet{ganji2020pitfalls}.  We numerically evaluate $d_f(0.1)$, the ratio
$L(n,d_f)/2^n$, and the smallest set accumulating $90\%$ Fourier mass for
$n=10,\ldots,18$ and $x\in\{1,2,4,6,8\}$.  Distinct-character recovery is
measured by repeated exact Fourier sampling until $50\%$ and $90\%$ of that
set has appeared.

\subsection{Access-matched algorithms}

\paragraph{Ideal Fourier sampler.}
Starting from $2^{-n/2}\sum_u\lvert u\rangle$, applying $O_f$ and a final
Hadamard layer measures $S$ with probability $\widehat f(S)^2$.  This is an
ideal query experiment: its timing excludes oracle synthesis, fault-tolerant
overhead, and state preparation.  Quantum example-oracle learning has long
made this access distinction explicit \citep{bshouty1998dnf}.

\paragraph{Classical comparator.}
KM recursively estimates the Fourier energy below a prefix and prunes prefixes
whose estimated energy lies below a threshold \citep{kushilevitz1993learning}.
Like the related Goldreich--Levin heavy-coefficient method
\citep{goldreich1989hardcore}, it requires membership queries but not an
enumerated truth table or an exact spectrum.  Our implementation caches
distinct model evaluations; exact spectra are used only to score recovered
characters.  We report both distinct model evaluations and logical estimator
lookups and do not translate them into coherent-oracle calls.  The comparison
uses $n\in\{10,12,14\}$, five instances each of APUF, 2-XOR, and 4-XOR,
8,192 KM energy samples, and 64 ideal-sampling trials.  We score recovery of
characters with power $\ge0.01$ (threshold-heavy characters) and of the
smallest exact sets carrying $50\%$ and $90\%$ mass; each
ideal trial receives $2^n$ calls.  This deliberately generous quantum-side
budget reaches full truth-table cardinality; it is a stress test of the sampler,
not a proposed practical query budget.

\paragraph{Kernel diagnostic.}
As a separate route to quantum learning, we test a two-layer, degree-2,
scale-0.6 parity-aligned quantum feature map (PA-QFM), following the quantum
kernel setting of \citet{havlicek2019supervised}.  Before fitting labels, we
compare classical and quantum Gram matrices $K_C$ and $K_Q$ using
Huang et al.'s asymmetric geometric difference
\begin{equation}
g(K_C\Vert K_Q)=
\sqrt{\left\|\sqrt{K_Q}K_C^{-1}\sqrt{K_Q}\right\|_\infty},
\label{eq:gcq}
\end{equation}
after trace normalization \citep{huang2021power}; we denote this quantity by
$g_{CQ}$.  It implies
$s_{K_C}\le g^2s_{K_Q}$ for the corresponding model-complexity measure.  We
minimize $g$ label-independently over efficient structured-parity,
suffix-RBF, and suffix-polynomial kernels.  Exact support mismatch is reported
as infinite rather than hidden by a pseudoinverse; ridge sensitivity is a
separate diagnostic.  We then evaluate Huang et al.'s label-specific measure
\begin{equation}
s_K=y^T K^{-1}y
\label{eq:sk}
\end{equation}
on the same five $n=16,N=512$ pools.  For each of three deterministic PUF
instances per pool and each construction (45 label tasks), the classical side
minimizes exact $s_C$ over the suite fixed before evaluation.  Under the bound, $s_C/s_Q>1$
favors quantum label alignment.  This is an in-sample complexity diagnostic,
not held-out accuracy.  To control conditioning and model selection, we apply
128 deterministic permutations to each of the 15 4-XOR label vectors while
preserving class balance, pools, Gram matrices, and the complete kernel suite.
The predeclared primary endpoint is the upper-tail plus-one permutation test of
the median zero-ridge ratio.  Symmetric trace-normalized ridge values
$\lambda\in\{10^{-4},10^{-3},10^{-2},0.1,1\}$ are sensitivity analyses.

\subsection{A reversible sign oracle and approximation criteria}
\label{sec:oracle}

At fractional precision $p$, let $q_i=\operatorname{round}(2^pw_i)$.  In the
suffix basis, the quantized delay is
\begin{equation}
D_q(u)=q_b+\sum_iq_i(1-2u_i)=K+\sum_i(-2q_i)u_i.
\label{eq:delay}
\end{equation}
We choose a two's-complement accumulator width $m$ satisfying
$2^{m-1}-1\ge |q_b|+\sum_i|q_i|$, preventing signed wraparound.  A QFT
constant-adder chain \citep{draper2000addition} accumulates Eq.~\eqref{eq:delay};
the most significant bit controls a $\lvert-\rangle$ kickback qubit; and the
accumulator and kickback preparation are fully inverted.  Exhaustive
basis-state tests verify the signed phase and clean ancillas, while a final
Hadamard layer is compared with the exact Walsh spectrum.

We compare the full QFT with an exact nearest-neighbor line-QFT construction
after optimization-level-3 transpilation to the fixed
\texttt{FakeSherbrooke} coupling/calibration snapshot in Qiskit 2.5.2
\citep{javadiabhari2024qiskit}.  No quantum processing unit (QPU) or Runtime
execution is performed.
We fix two separate approximation criteria before the precision sweep:

\begin{itemize}
  \item a finite audit gate: worst-instance Fourier-distribution total
  variation (TV) $\le0.01$ and exact equality of all characters having power
  at least $0.01$ in either floating or quantized model;
  \item a model-independent certificate: phase-state trace-distance upper
  bound $\sqrt{1-(1-2\eta)^2}\le0.10$, where $\eta$ is truth-table response
  disagreement.
\end{itemize}

These gates are reported separately.  Duration divided by mapped-qubit median
$T_2$ is a coherence warning, not a success probability.

\paragraph{Protocol 1 (apply before training or hardware use).}
Pre-register a Fourier mass target $\rho$, maximum auditable support $B_H$,
classical and coherent-query budgets $(B_C,B_Q)$, the prediction-advantage
factor being claimed, task-error limits, and a target-device resource budget.
Then apply the gates in order:
\begin{enumerate}
  \item \emph{Structure passes} only if both the degree region
  $L(n,d_f(1-\rho))$ and the smallest $\rho$-mass set have cardinality at most
  $B_H$ on the declared instance distribution.
  \item \emph{Algorithmics passes} for Fourier recovery only if the quantum
  sampler reaches $\rho$ mass within $B_Q$ while every access-matched classical
  baseline exceeds $B_C$; a kernel claim additionally requires the measured
  $g_{CQ}^2$ bound to meet the factor actually claimed across predeclared pools,
  followed by a predeclared label-complexity ratio $s_C/s_Q$; prediction claims
  additionally require held-out task performance.
  \item \emph{Implementation passes} only if the oracle phase and uncomputation
  are exact on ground-truth tests and at least one precision meets both the
  task-error and target-device budgets.  A duration/$T_2$ ratio is a warning,
  not a substitute for a declared native-resource budget.
\end{enumerate}
Failure of a gate blocks an end-to-end advantage claim but does not invalidate
upstream measurements; each gate is therefore reported separately.

\section{Results}
\label{sec:results}

\subsection{Low degree is not small support at reachable \texorpdfstring{$n$}{n}}

\begin{figure}[t]
\centering
\includegraphics[width=\linewidth]{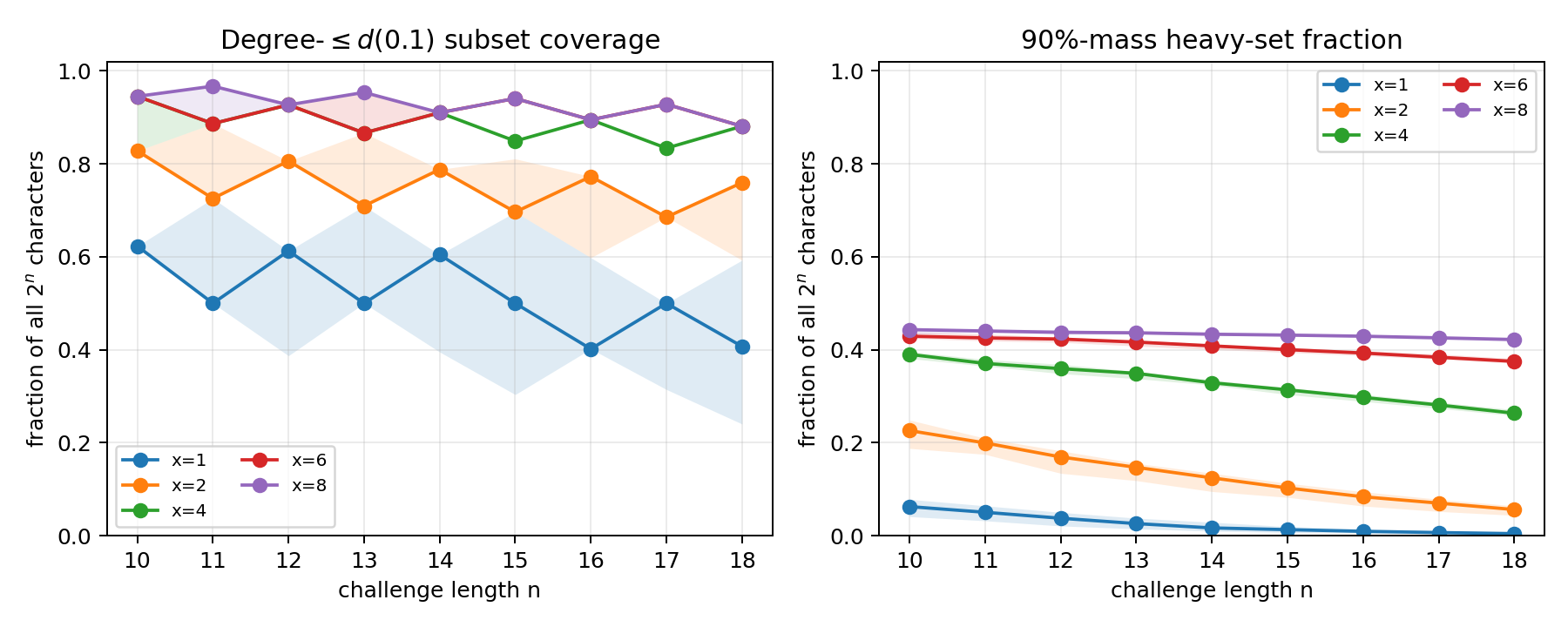}
\caption{Finite-$n$ concentration sweep with 30 deterministic seeded PUF
instances per cell.  Lines are medians and shaded bands are interquartile
ranges across instances.  Left: the fraction of all $2^n$ characters admitted
by degree $\le d_f(0.1)$.  Right: the fraction needed to accumulate $90\%$
Fourier mass.  For higher XOR counts, both denominators remain large rather
than shrinking toward a sparse audit target.}
\label{fig:concentration}
\end{figure}

Figure~\ref{fig:concentration} shows the denominator that a degree-only claim
hides.  Across 30 seeded instances at $n=14$, 4-XOR has median
$d_f(0.1)=9$ (interquartile range, IQR 9--9), but degree at most 9 includes 14,913 of 16,384
characters ($91.02\%$); the median $90\%$-mass set occupies $32.90\%$ of the
spectrum (IQR $32.41$--$33.52\%$).  At $n=18$, median degree-support fraction
is $88.11\%$ for both 4-XOR and 8-XOR.  Their median $90\%$-mass sets contain
69,075 ($26.35\%$, IQR $25.84$--$27.10\%$) and 110,641 ($42.21\%$, IQR
$41.84$--$42.41\%$) characters, respectively.  Median recovery of $90\%$ of
these distinct sets takes 596,793 and 507,352 ideal Fourier samples.  The
nonmonotone count reflects unequal spectral masses, but both cases pay
heavy-set cardinality.  Proposition~\ref{prop:xor} is
asymptotically valid; at our parameters the scale $x^2/\varepsilon^2$ already
exceeds the Boolean cube's maximum possible Fourier degree $n$, and therefore
supplies no useful finite-$n$ sparsity certificate.

\subsection{Access-matched threshold and mass recovery}

We evaluate 45 rows at power threshold $0.01$.  KM recovers every
threshold-heavy character in all 44 nonempty rows; one 4-XOR row is empty.
Yet Table~\ref{tab:mass-audit} shows why this is not a strong benchmark: the
median thresholded set has only four characters for 4-XOR and carries $6.5\%$
of its spectrum.  Every KM row queries all $2^n$ challenges at least once
(1,024--16,384 distinct evaluations) and uses 16,384--245,760 logical
estimator lookups, so this sweep demonstrates neither exponential enumeration
nor sub-domain KM savings.

\begin{table}[t]
\caption{Threshold and Fourier-mass recovery, medians over 15 rows per
construction.  ``QFS mass'' uses the fixed $2^n$ ideal-call budget;
Table~\ref{tab:audit-dispersion} gives interquartile ranges.
$H_{.01}$ contains characters with power at least $0.01$;
$H_{.9}$ is the smallest set carrying $90\%$ mass.}
\label{tab:mass-audit}
\centering
\small
\begin{tabular}{lrrrrrr}
\toprule
 & $|H_{.01}|$ & mass$(H_{.01})$ & KM mass & QFS mass & $|H_{.9}|$ & KM recall $H_{.9}$ \\
\midrule
APUF  & 11 & .727 & .735 & .935 & 89    & .157 \\
2-XOR & 15 & .403 & .445 & .890 & 716   & .032 \\
4-XOR & 4  & .065 & .123 & .832 & 1,415 & .0065 \\
\bottomrule
\end{tabular}
\end{table}

The ideal sampler also recovers every threshold-heavy set, with median full-set
first passage 70.5--253 calls.  Mass-relative recovery reverses the impression:
pooled across all 45 rows, the median $90\%$-mass set has 400 characters
(maximum 5,794),
KM's median recall of that set is $4.1\%$, and median ideal first passage among
rows that reach $90\%$ mass is 2,041 calls.  No 4-XOR row reaches $90\%$ mass
within its $2^n$ budget.  Thus spectral diffuseness, not merely the choice of
classical algorithm, is the binding obstacle.  Classical evaluations and
coherent oracle calls remain different resources; this result does not
combine them into a single cost or prove equality of query complexity.

\subsection{Kernel diagnostics under conditioning and permutation controls}

The kernel diagnostic initially appears more favorable.  At $n=16$, the
median best $g_{CQ}$ rises from $1.118$ at $N=32$ to $1.280$ at $N=128$,
$1.871$ at $N=256$, and $2.151$ at $N=512$.  The five $N=512$ pools span
$2.028$--$2.179$, giving maximum $g_{CQ}^2=4.749$.  A descriptive fit over the
six measured sizes is $g_{CQ}=0.396N^{0.265}$ ($R^2=0.890$ in log space), with
no observed plateau through $N=512$; we do not extrapolate this fit.  At
$N=512$, quantum Gram matrices remain close to identity (median
effective-rank fraction $0.999457$ and mean off-diagonal
$1.07\times10^{-4}$), whereas the selected suffix-RBF classical matrices do
not (median $0.464068$ and $2.99\times10^{-2}$, respectively).  The increase
reflects different Gram geometries rather than two near-identity kernels, but
it is substantially conditioning-driven.  Indeed, when $K_Q\approx I$ after
trace normalization, Eq.~\eqref{eq:gcq} gives
$g_{CQ}^2\approx\lVert K_C^{-1}\rVert_\infty
=1/\lambda_{\min}(K_C)$.  Across the five pools,
$\operatorname{corr}(g_{CQ},1/\sqrt{\lambda_{\min}(K_C)})=0.991$, and median
$\lambda_{\min}(K_C)=0.21584$ gives $1/\sqrt{\lambda_{\min}}=2.15244$.  Symmetric
ridge reduces median $g_{CQ}$ to $1.866$ at $\lambda=0.1$ and $1.281$ at
$\lambda=1$.  Because $g_{CQ}$ is a label-independent upper bound, geometry
alone does not establish prediction advantage.  Across 15 rows per construction,
median $s_C/s_Q$ is $0.608$ for APUF, $0.995$ for 2-XOR, and $1.156$ for
4-XOR; the counts above one are 0/15, 7/15, and 15/15.  The 4-XOR range is
$1.085$--$1.233$, with degree-6 structured parity the best classical kernel in
every row.  The permutation control reverses its interpretation: the joint
null median is $1.176$, exceeding the observed $1.156$, with upper-tail
$p=0.930$.  All 1,920 null fits also select degree-6 structured parity, and the
ridge-grid $p$-values remain $0.930$--$0.977$.  We therefore find no evidence
that the above-one ratio is specific to the 4-XOR labels.
Table~\ref{tab:gate-summary} summarizes what each gate establishes.

\begin{table}[t]
\caption{Decision-gate summary.  ``No'' means unsupported in the measured
regime, not a universal impossibility theorem.}
\label{tab:gate-summary}
\centering
\small
\begin{tabular}{p{0.20\linewidth}p{0.47\linewidth}p{0.19\linewidth}}
\toprule
Gate & Evidence & Verdict \\
\midrule
Structure & 4-XOR at $n=14$: median $d=9$ covers $91.02\%$; heavy set is $32.90\%$ & No \\
Algorithmic: Fourier & $|H_{.9}|$ median 400; KM recall $4.1\%$; 4-XOR QFS never reaches $90\%$ mass & No cheap recovery \\
Algorithmic: kernel & $g_{CQ}$ tracks conditioning ($r=.991$); 4-XOR permutation $p=.930$ & No label-specific evidence \\
Implementation & exact oracle; estimated durations at certified precisions exceed median $T_2$ & Snapshot warning; no QPU execution \\
\bottomrule
\end{tabular}
\end{table}

\subsection{Oracle precision and cost on the target snapshot}

Across 81 matched $(n,p,\text{instance},\text{seed})$ cases for
$n\in\{8,10,12\}$ and $p\in\{4,6,8\}$, the exact line-QFT reduces median
routed native-gate depth by $18.9\%$, estimated duration by $19.2\%$, and two-qubit
count by $8.2\%$ relative to the full QFT.  It wins 74/81 depth comparisons
and 73/81 duration comparisons, so selection remains layout-dependent.  An
independently validated weighted-comparator alternative is decisively worse in
the matched $n=8,p=4$ probe: depth 25,908, 8,613 two-qubit gates, 32 logical
qubits, and duration $16.85T_2$, versus line-QFT depth 1,555, 850 two-qubit
gates, 17 qubits, and $0.90T_2$.

\begin{figure}[t]
\centering
\includegraphics[width=\linewidth]{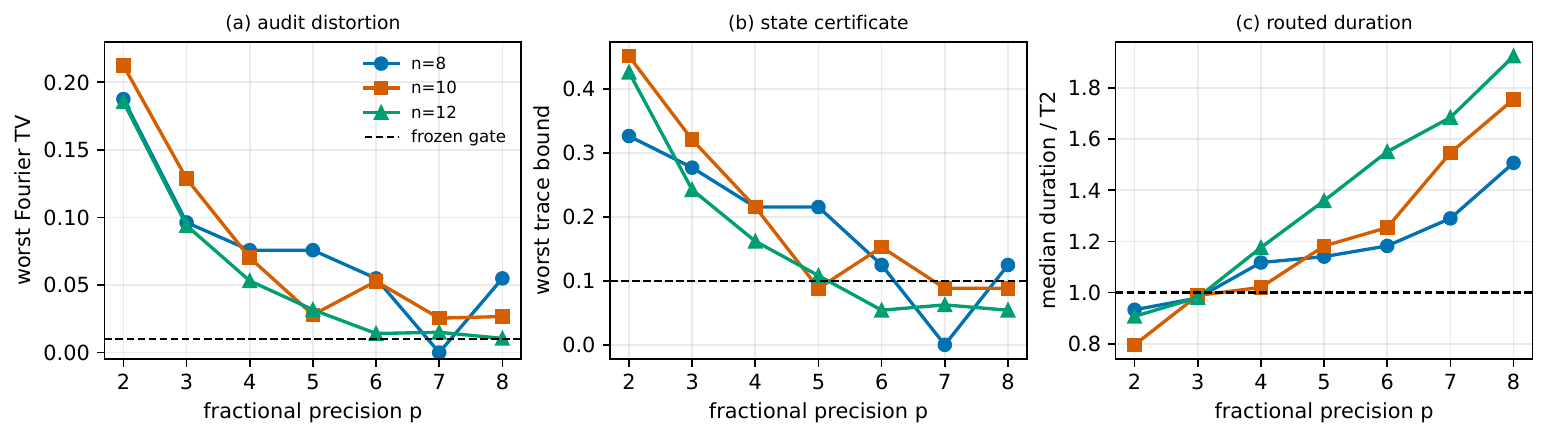}
\caption{Precision and resource tradeoffs over three PUF instances and three
transpiler seeds per point; accuracy is worst-instance and routed cost is the
median.  ``Frozen gate'' denotes a criterion fixed before the sweep.  Dashed
lines show these thresholds.  Signal-preserving
precision arrives only after median circuit duration reaches or exceeds mapped
median $T_2$.  Transpilation only; no backend execution.}
\label{fig:precision}
\end{figure}

Figure~\ref{fig:precision} gives the task-aware precision sweep.  Low precision
can fit below median $T_2$ but distorts the target spectrum.  The least
model-independent precisions are $p=7,5,6$ at $n=8,10,12$; their line-QFT
depths are 2,028, 1,940, and 2,340, and their median durations are
$1.29T_2$, $1.18T_2$, and $1.55T_2$.  Only $(n,p)=(8,7)$ passes the finite
audit gate.  No tested precision passes it at $n=10$ or 12; even $n=12,p=8$
has worst TV $0.0105$.  Accuracy is nonmonotone because rounding can move a
near-zero delay margin across the sign boundary.  There is therefore no
tested precision that meets both the stated accuracy criteria and the median
coherence-time threshold.  This is a feasibility screen for
the declared snapshot, not a measured device failure rate or a bound covering
other hardware.

\paragraph{Fault-tolerant logical estimate.}
The coherence-time comparison does not decide whether this oracle is useful on a
future error-corrected machine.  We therefore inspect the exact phase angles of
the validated line-QFT circuit before routing.  Each controlled phase is
expanded into three single-qubit phase rotations and two CNOTs, without
cross-gate cancellation.  Clifford and exact-$T$ angles are separated from
arbitrary rotations.  For the latter, Table~\ref{tab:logical-resources} assigns
a uniform per-rotation error whose triangle-inequality sum is at most the
declared whole-oracle budget.  We report both the typical Ross--Selinger leading
term and Selinger's constructive $K+4\log_2(1/\epsilon)$ estimate with $K=10$
\citep{ross2016optimal,selinger2015efficient}.

\begin{table}[t]
\caption{Logical resources at the least model-independently certified
precisions, over three PUF instances. $T$ columns are median constructive
estimates for whole-oracle synthesis error $\delta$.}
\label{tab:logical-resources}
\centering
\small
\begin{tabular}{rrrrrrr}
\toprule
$n$ & $p$ & qubits & arbitrary $R_z$ & CNOT & $T$, $\delta=10^{-3}$ & $T$, $\delta=10^{-6}$ \\
\midrule
8  & 7 & 20--21 & 806--1,074 & 985--1,177 & 79,550 & 113,948 \\
10 & 5 & 20--21 & 630--814   & 769--925   & 69,768 & 101,048 \\
12 & 6 & 24     & 1,046--1,072 & 1,165--1,185 & 94,514 & 136,434 \\
\bottomrule
\end{tabular}
\end{table}

These are per-oracle-call constructive logical estimates, not optimized lower
bounds, physical gate counts, or a fault-tolerant runtime.  A
surface-code cycle estimate additionally requires a physical error rate, code
distance, decoder, magic-state factory, connectivity, and cycle time.  Rather
than select those assumptions to force a conclusion, we expose the threshold:
even the smallest evaluated oracle requires about $7\times10^4$ constructive
$T$ gates at $\delta=10^{-3}$ before repeated Fourier samples are counted.

\section{Discussion}

\paragraph{Interpreting the controlled comparisons.}
Low-degree concentration plus ideal Fourier sampling does not establish a
realizable quantum advantage for these designer-held delay models.  The
structural support is broad at reachable $n$; threshold recovery captures little
mass for 4-XOR
while mass recovery is expensive for both KM and ideal sampling; access
matching invalidates brute-force support enumeration as the classical
baseline; and the tested signal-preserving oracle precisions exceed the
snapshot's coherence scale in estimated duration.  The kernel gate also fails
after controls: its
geometric difference grows with an ill-conditioned classical direction, and
the above-one 4-XOR label ratios are typical under balance-preserving
permutations.  This is more informative than a classifier accuracy tie because
it identifies why two analytic diagnostics appear favorable before training
yet do not isolate task-specific quantum alignment.

\paragraph{Methodological implication beyond PUFs.}
Whenever a trace-normalized quantum Gram matrix is near identity while a
classical Gram develops a small eigenvalue, $g_{CQ}$ can grow through the
inverse in Eq.~\eqref{eq:gcq} without corresponding task-label alignment.
Geometric-difference diagnostics should therefore report
$\lambda_{\min}(K_C)$, ridge sensitivity, and a label-permutation control rather than treating
$g_{CQ}$ alone as evidence of useful quantum geometry.

\paragraph{What it does not say.}
We do not prove that every classical algorithm matches Fourier sampling query
for query, nor that every quantum feature map is classically competitive.
Indeed, the ideal sampler uses fewer coherent calls on the threshold-heavy
task than KM uses distinct model evaluations in this finite sweep.
Fault-tolerant hardware, a lower-depth
sign oracle, or a distribution with a genuinely sparse spectrum could change
the Fourier-audit conclusion.  A different quantum feature map, classical
suite, or label distribution could still pass the kernel gate; our null result
is specific to the declared PA-QFM and simulated PUF family.  Each possibility
can be assessed using the protocol's explicit criteria.

\paragraph{Extending the protocol.}
Other Boolean PUF models require a declared challenge distribution and
representation, measured spectral support, access-matched classical baselines,
and a separately validated reversible response oracle.  The XOR concentration
argument does not automatically extend to other constructions.  For audits
beyond Fourier recovery, the structural diagnostic and success criterion must
be derived from the new task.  Access matching and implementation accounting
remain applicable; our numerical verdicts do not transfer without evaluation.

\paragraph{Security interpretation.}
The holder synthesizing $O_f$ already possesses the model: this is a
designer-side diagnostic, not a new capability for a remote attacker.  The
benefit is a reproducible screen against unsupported quantum security claims.
The dual-use risk is that audit code can also assist classical PUF modeling
when a delay model is available; it does not create coherent access to silicon.
We keep access classes explicit and publish resource estimates with the code.

\section{Limitations}
\label{sec:limitations}

The experiments use additive-delay APUF and XOR-APUF simulations under uniform
challenges.  Agreement with \texttt{pypuf} catches implementation errors but is
not silicon validation; inter-die process and environmental behavior are
outside the central experiments.  The exact concentration sweep stops at
$n=18$, the KM study at $n=14$, and exact quantum kernels at $n=16,N=512$.
Thirty deterministic instances per concentration cell quantify seed-to-seed
dispersion but do not estimate a fabricated-device population.  The geometric
diagnostic covers five pools and a declared efficient classical suite, not all
kernels or label distributions.
The label-complexity comparison adds 45 deterministic tasks on those same pools; selecting the
minimum $s_C$ over the suite is an in-sample complexity comparison, not model
selection on an independent validation split or a held-out accuracy estimate.
The permutation control uses 128 draws per row (minimum attainable plus-one
$p=1/129$) and tests label specificity conditional on the observed pools and
kernel suite; it is not a population-level generalization experiment.

The KM implementation exhausts the finite domains, so we do not empirically
claim its asymptotic query efficiency.  Mass-relative target sets are defined
using exact spectra for scoring only; neither algorithm receives them.
Conversely, ideal Fourier timing omits
fault-tolerant and oracle costs.  \texttt{FakeSherbrooke} is a static
calibration/coupling snapshot; duration/$T_2$ is an optimistic warning rather
than a noise simulation or fidelity prediction.  The precision experiment
uses only three simulated PUF instances per point.  No QPU was executed, and no
$n=32$--64 tensor-network kernel result is established.  Accordingly, our
conclusion is a boundary for the measured models and access assumptions, not a
general lower bound on quantum learning.

Post-silicon validation would require measured CRPs with disjoint model-fitting
and evaluation challenges, held-out devices for population claims, and repeated
measurements across operating conditions.  A future study should separate
response noise from model mismatch and test the audit's stability under the
measured uncertainty.  This would validate the model being audited, not provide
coherent access to a fabricated PUF.

The logical Clifford+$T$ estimate uses a uniform synthesis-error allocation
and does not optimize rotations jointly.  It is an explicit,
architecture-independent accounting point, not a surface-code resource
estimate or a forecast of fault-tolerant feasibility.  The geometric
difference is an upper bound optimized over label vectors and is sensitive to
small eigenvalues.  The 4-XOR median $s_C/s_Q$ does not exceed its balance-preserving
null and therefore supplies no label-specific positive result.

\section{Conclusion}

An ideal quantum query advantage is not yet an application advantage.  For
delay-PUF model auditing, none of the three criteria establishes a practical
advantage in the measured regime: spectra
are not sparse at reachable $n$; the kernel diagnostics do not survive
conditioning and permutation controls; and signal-preserving reversible
arithmetic exceeds the snapshot's coherence scale in estimated duration despite
an exact depth reduction.  Coherent access also remains designer-only, with KM as the access-matched
classical baseline.  Reporting the diagnosed negative outcome makes the
Three-Gate Quantum Audit Protocol a reusable
evaluation tool for secure and trustworthy QML.  Future positive results
should state which access class they inhabit, carry concentration denominators,
connect kernel geometry to task labels, and include the oracle in their
physical resource budget.

\begin{ack}
The computation for this work was performed on the high performance computing infrastructure provided by Research Support Services at the University of Missouri, Columbia MO. DOI: 10.32469/10355/97710
\end{ack}

\bibliographystyle{plainnat}
\bibliography{references}

\appendix

\section{Proof of Proposition~\ref{prop:xor}}
\label{app:proof}

Couple $u$ and $u_\delta$ by independently flipping every input bit with
probability $\delta$.  Because all factors are $\{-1,+1\}$-valued, a product
change implies that at least one factor changed.  Pointwise,
\[
\mathbf{1}\!\left[\prod_{k=1}^x A_k(u)\ne
\prod_{k=1}^x A_k(u_\delta)\right]
\le \sum_{k=1}^x\mathbf{1}[A_k(u)\ne A_k(u_\delta)].
\]
Taking expectations and applying the noise-stability bound for weighted
majority gives
\begin{equation}
\NS_\delta(f)\le\sum_{k=1}^{x}\NS_\delta(A_k)
\le Cx\sqrt{\delta}
\label{eq:nsunion}
\end{equation}
for a universal constant $C$ \citep{peres2004noise}.  For a Boolean function,
\begin{equation}
\NS_\delta(f)=\frac12\sum_{S\subseteq[n]}
\left[1-(1-2\delta)^{|S|}\right]\widehat f(S)^2.
\label{eq:nsfourier}
\end{equation}
Choose $\delta=1/(2d)$ for $d\ge1$.  For $|S|>d$,
$1-(1-2\delta)^{|S|}\ge1-(1-1/d)^d\ge1-e^{-1}$.
Equations~\eqref{eq:nsunion}--\eqref{eq:nsfourier} then imply
\[
\sum_{|S|>d}\widehat f(S)^2
\le \frac{2C}{1-e^{-1}}\frac{x}{\sqrt{2d}}
=O\!\left(\frac{x}{\sqrt d}\right).
\]
Setting the right-hand side to $\varepsilon$ yields
$d=O(x^2/\varepsilon^2)$.  When $d$ is fixed relative to $n$,
$\sum_{j\le d}\binom{n}{j}=n^{O(d)}$, giving the stated formal support bound.
No clipping operation is used, so the degree conclusion is preserved.

\section{Oracle derivation and validation}
\label{app:oracle}

Equation~\eqref{eq:delay} uses the challenge register directly because the
audit is performed in the suffix-parity basis.  With accumulator modulus
$N=2^m$, QFT addition multiplies Fourier basis state $\lvert y\rangle$ by
$\exp(2\pi iay/N)$ before applying the inverse QFT.  Factoring
$y=\sum_j2^jy_j$ gives one phase rotation of angle $2\pi a2^j/N$ per
accumulator qubit; controlling those rotations on $u_i$ adds $-2q_i$ exactly
when $u_i=1$.  The accumulator-width bound prevents modular overflow from
changing the sign.  A CNOT from the most significant bit to
$\lvert-\rangle$ contributes $-1$ exactly when $D_q(u)<0$.  Inverting all
arithmetic returns the accumulator and kickback ancilla to zero.

Tests exhaust every basis state for small signed examples, compare the phase
of every challenge with the quantized model, assert clean ancillas, cover
constant-response and overflow edge cases, and compare the final probability
distribution with an exact Walsh spectrum.  Hardware-facing code is admitted
only after these simulator tests pass.  The weighted-comparator alternative is
tested with statevectors only through $n=3$ because its workspace grows the
Hilbert space rapidly; the reported $n=8$ result is transpilation only.

\section{Experimental protocol and provenance}
\label{app:protocol}

Every entry point takes an explicit seed.  Each result JSON records the full
argument vector, Git SHA, dirty-worktree flag, UTC timestamp, seeds, and package
versions.  Table~\ref{tab:provenance} maps manuscript results to their committed
generators.  Challenge pools are sampled without replacement.  Any
sample-complexity experiment enforces
$\max(N_{\rm train})+N_{\rm test}\le2^n/100$; the default grid therefore
requires $n\ge20$.

Source code, evidence JSONs, pinned requirements, reproduction commands, and a
deterministic hash manifest are available in an
\href{https://figshare.com/s/db4565924150416f9788}{anonymous reproduction archive}.

\begin{table}[ht]
\caption{Machine-readable evidence used in the paper.}
\label{tab:provenance}
\centering
\small
\begin{tabular}{p{0.26\linewidth}p{0.30\linewidth}p{0.30\linewidth}}
\toprule
Result & Generator & Evidence JSON \\
\midrule
Concentration & \path{sweep_concentration.py} & \path{phase0_concentration_sweep.json} \\
KM versus QFS & \path{benchmark_fourier_audits.py} & \path{phase1_findings.json} \\
Kernel geometry & \path{measure_geometric_difference.py} & \path{phase1_findings.json} \\
Kernel labels & \path{measure_label_kernel_complexity.py} & \path{phase1_findings.json} \\
Oracle comparison & \path{compare_e2_oracles.py} & \path{e2_oracle_comparison_findings.json} \\
Weighted comparator & \path{compare_e2_oracles.py} & \path{weighted_probe.json} \\
Precision frontier & \path{summarize_e2_precision_frontier.py} & \path{e2_precision_frontier_findings.json} \\
Logical oracle cost & \path{estimate_e2_logical_resources.py} & \path{e2_logical_resources.json} \\
\bottomrule
\end{tabular}
\end{table}

\begin{table}[ht]
\caption{Interquartile ranges corresponding to the construction-level medians
in Table~\ref{tab:mass-audit}.  Each range is descriptive dispersion over 15
rows spanning $n\in\{10,12,14\}$ and five PUF instances per $n$, not a
population confidence interval.}
\label{tab:audit-dispersion}
\centering
\scriptsize
\begin{tabular}{lrrrrrr}
\toprule
 & $|H_{.01}|$ & mass$(H_{.01})$ & KM mass & QFS mass & $|H_{.9}|$ & KM recall $H_{.9}$ \\
\midrule
APUF  & 11--13 & .687--.748 & .715--.769 & .928--.946 & 74.5--182 & .069--.195 \\
2-XOR & 12--18.5 & .385--.453 & .414--.508 & .876--.901 & 272--1,672.5 & .011--.098 \\
4-XOR & 3.5--7.5 & .044--.124 & .076--.177 & .825--.841 & 406.5--4,732 & .0020--.038 \\
\bottomrule
\end{tabular}
\end{table}

The precision-frontier Slurm job used one task, two CPUs, a 32-GB allocation,
and a two-hour limit, with a 30-minute timeout for each $n$ stage.  It completed
within the allocation, but peak memory was not preserved in the immutable job
artifact and is therefore not reported.  The exact run covered three PUF
instances, seven precisions, two exact oracle designs, and three transpiler
seeds.  The mass-audit and kernel-geometry follow-up used the same CPU and
memory allocation with a 2.5-hour limit; it completed in 77 seconds with
43,708 KiB peak resident memory.  Other reported sweeps are CPU-only and retain
per-run wall times in their JSONs; failed bounded tensor-network and arithmetic probes are preserved
rather than omitted from the artifact.  No GPU or QPU was used for the results
in this paper.

The released code is MIT-licensed.  The externally used software assets are
Qiskit 2.5.2 and Qiskit IBM Runtime 0.49.0 under Apache 2.0, and
\texttt{pypuf} 2.2.0 under GPLv3; versions are also recorded in provenance.

\end{document}